\documentstyle[twocolumn,floats,prl,aps,epsf,psfrag]{revtex}
\begin{document}
\draft

\twocolumn[\hsize\textwidth\columnwidth\hsize\csname@twocolumnfalse\endcsname

\title{Perfect quantum error correction coding in 24 laser pulses}
\author{Samuel L.~Braunstein}
\address{Universit\"at Ulm, Abteilung Quantenphysik, 89069 Ulm, Germany}
\address{\cite{CurrAd}SEECS, University of Wales, Bangor, Gwynedd LL57 1UT, UK}
\author{John A.~Smolin}
\address{IBM Research Division, Yorktown Heights, NY 10598}
\date{\today}

\maketitle

\begin{abstract}
An efficient coding circuit is given for the perfect quantum error 
correction of a single qubit against arbitrary 1-qubit errors within 
a 5 qubit code. The circuit presented employs a double `classical' code, 
i.e., one for bit flips and one for phase shifts.  An implementation of 
this coding circuit on an ion-trap quantum computer is described that 
requires 26 laser pulses. A further circuit is presented
requiring only 24 laser pulses, making it an efficient protection scheme
against arbitrary 1-qubit errors. In addition, the performance of two 
error correction schemes, one based on the quantum Zeno effect and the 
other using standard methods, is compared. The quantum Zeno error 
correction scheme is found to fail completely for a model of noise based 
on phase-diffusion.
\end{abstract}


\vskip2pc]

Quantum error correction schemes 
\cite{Shor,Shor2,Steane,smolin,Zurek,Steane2,Ekert,newpaper} 
hold the promise of reliable storage, processing and transfer of
quantum information. They actively `isolate' a quantum system from 
perturbations, which would otherwise decohere the state 
\cite{Unruh,Landauer}. How quickly this decoherence occurs depends to 
a large extent on what degrees of freedom are involved: single- or 
many-body, electronic, nuclear, etc. In principle, however, the 
development of quantum error correction allows one to decouple a quantum 
state from arbitrary few-particle perturbations. 

The decoupling in quantum error correction schemes is achieved by
unitarily `rotating' the state into one involving a larger
number of degrees of freedom. In this larger space the information 
about the original state is recorded only in multi-particle 
correlations. Thus, if only a few particles undergo decohering
perturbations, the multi-particle correlations are not destroyed, but
only mixed amongst each other. After determining which few particle 
perturbation has occurred we can unmix the multi-particle correlations 
and hence reconstruct the original state. If, by contrast, decohering 
perturbations accumulate over too many particles then the 
multi-particle correlations are no longer isolated and the error 
correction begins to break down. 

In this paper an efficient coding circuit for arbitrary single-qubit 
errors is given. Its efficiency is quantified relative to a specific 
quantum computer model --- Cirac and Zoller's ion-trap model \cite{CZ}. 
Next, two schemes designed to protect against single-qubit phase-noise
are studied. One scheme relies on the quantum Zeno effect \cite{Ch,V} and
uses two qubits to protect against `slow' perturbations of the system;
the other is a more conventional quantum error correction scheme 
\cite{Steane2,Ekert,ME} that requires three qubits to protect 
against arbitrary single-particle dephasing. The poor behavior
of the Zeno schemes is discussed and explained.

\section*{Efficient coding}

Various authors \cite{smolin,Zurek,newpaper} have presented circuits
implementing a 5-qubit which protects one qubit of quantum
information.  This code is described as `perfect' since it allows for
the complete correction of arbitrary single qubit errors. (The term
qubit \cite{Schu} represents the amount of `quantum' information
stored in an arbitrary two-state quantum system.)  In this section a
simpler version of the Laflamme {\it et al} coding circuit
\cite{Zurek} is presented. We discuss the structure of the circuit and
consider its efficiency. The measure of efficiency used
\cite{Preskill} is the number of laser pulses required to implement
the scheme on an ion-trap quantum computer. A second circuit yielding
a slightly different version of this code was found by a computer
search and is the most efficient circuit so far constructed for
one-bit encoding.

\begin{figure}[thb]
\begin{psfrags}
\psfrag{p1}[c]{\Large $~~|\psi\rangle$}
\psfrag{z}[c]{\Large $|0\rangle$}
\psfrag{u1}[cb]{\small $~~~~\,\hat U$}
\psfrag{u2}[cb]{\small $~~~~~\hat U^\dagger$}
\epsfxsize=3.4in
\epsfbox[-40 -10 210 165]{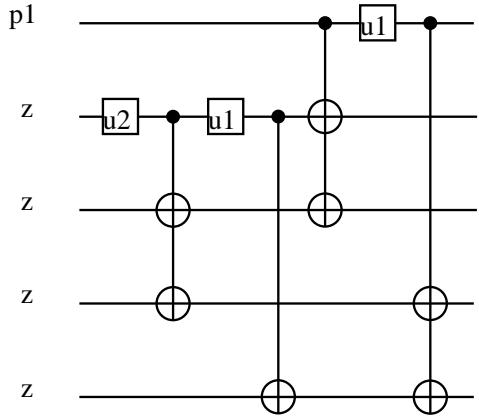}
\end{psfrags}
\caption{Efficient quantum 5-qubit error correction circuit. The 
system starts at the left and is successively processed through
each of the elementary gates proceeding from left to right. Here
the qubit $|\protect\psi\protect\rangle$ is rotated into a
protective 5-particle state by the unitary operations represented
by the elements of this circuit. The 3-qubit gates shown are simply
pairs of controlled-NOT gates.}
\label{fig1}
\end{figure}

Fig.~\ref{fig1} shows our simplification of the 5-bit coding circuit of 
Laflamme {\it et al\/} \cite{Zurek}. This circuit uses single particle 
rotations
\begin{equation}
\hat U = {1\over\sqrt{2}} 
\left(\begin{array}{rr}
  1 & -1 \\
  1 &  1
\end{array}\right) \;,
\end{equation}
represented by the square `one-qubit' gates in the circuit,
and two-particle controlled-NOT gates
\begin{equation}
\begin{picture}(10, 5)
  \put(-2,9){\circle*{2.5}}
  \put(-2,9){\thinlines\line(0,-1){18}}
  \makebox(-4,-9)[c]{\large$\oplus$}
\end{picture} =
\left(\begin{array}{cccc}
  1 & 0 & 0 & 0 \\
  0 & 1 & 0 & 0 \\
  0 & 0 & 0 & 1 \\
  0 & 0 & 1 & 0 
\end{array}\right) 
\equiv
\begin{picture}(20, 5)
  \unitlength=1pt
  \put(30,15){\circle*{2.5}}
  \put(30,15){\thinlines\line(0,-1){16}}
  \put(30,15){\thinlines\line(1,0){18}}
  \put(30,15){\thinlines\line(-1,0){18}}
  \put(10,-9){\makebox(0,0){\framebox(15,15)[c]{\small$\hat U^\dagger$}}}
  \put(30,-9){\makebox(0,0){\framebox(15,15)[c]{$\hat\sigma_z$}}}
  \put(50,-9){\makebox(0,0){\framebox(15,15)[c]{\small$\hat U$}}}
  \put(17.5,-9){\thinlines\line(1,0){5}}
  \put(37.5,-9){\thinlines\line(1,0){5}}
\end{picture} ~~~~~~~~~~~~~ \label{eq1} \;;
\end{equation}
here the $\oplus$ notation is chosen because of the equality of the
controlled-NOT operation and the mathematical exclusive-OR operation.
The conditional $\hat \sigma_z$ operation itself is given by
\begin{equation}
\begin{picture}(20, 5)
  \unitlength=1pt
  \put(30,15){\circle*{2.5}}
  \put(30,15){\thinlines\line(0,-1){16}}
  \put(30,-9){\makebox(0,0){\framebox(15,15)[c]{$\hat\sigma_z$}}}
\end{picture} ~~~~~~~~~
\equiv
\left(\begin{array}{cccr}
  1 & 0 & 0 & 0 \\
  0 & 1 & 0 & 0 \\
  0 & 0 & 1 & 0 \\
  0 & 0 & 0 & -1
\end{array}\right) \label{eq1a} \;.
\end{equation}
Here these elements are represented in the basis where 
$|0\rangle=({1\atop 0})$, $|1\rangle=({0\atop 1})$ and
\begin{eqnarray}
|00\rangle = (1, 0, 0, 0)^{\rm T} \;,\nonumber \\
|01\rangle = (0, 1, 0, 0)^{\rm T}\;,\nonumber \\
|10\rangle = (0, 0, 1, 0)^{\rm T}\;,\\
|11\rangle = (0, 0, 0, 1)^{\rm T}\;,\nonumber 
\end{eqnarray}
etc. Decoding is executed by running the coding circuitry backwards, 
finally recovering the original state after a few extra operations 
\cite{Zurek,fn1}.

The circuit in Fig.~\ref{fig1} has an interesting structure:
The coding initially entangles four auxiliary particles and then
executes a double `classical' code on the state to be protected.
This `stores' the degrees of freedom of $|\psi\rangle$ in
the correlations of a five particle entangled state. The classical 
codes each consist of a simple triple redundancy of the qubit on the 
upper line of Fig.~\ref{fig1}: The first classical code may be interpreted
as protecting against random bit-flips and the second ---
against random phase shifts. This double code was motivated by theorem~6
of Steane \cite{Steane2}. He found, however, that such double coding
required a minimum of 7~qubits for a {\it linear\/} quantum code. The 
circuit in Fig.~\ref{fig1} produces a code that is not linear \cite{Zurek}.

To see that the above circuit reproduces the Laflamme {\it et al\/} code
it is sufficient to consider how this circuit acts on an arbitrary
qubit $|\psi\rangle=\alpha|0\rangle+\beta|1\rangle$. With the auxiliary
qubits initially in the states $|0\rangle$ the circuit generates
the superposition
\begin{eqnarray}
&&\phantom{+}\alpha\left(| 00000\rangle + | 00110 \rangle +| 01001 \rangle
- | 01111 \rangle \right.\nonumber \\
&&~~~\,\left.+ | 10011 \rangle+ | 10101 \rangle + | 11010 \rangle 
- | 11100 \rangle \right) \nonumber \\
&&+\beta \left(| 00011 \rangle- | 00101 \rangle- | 01010 \rangle
- | 01100 \rangle\right.\nonumber \\
&&~~~\,\left. -| 10000 \rangle + | 10110 \rangle +| 11001 \rangle 
+ | 11111 \rangle\right) \\
&=& \phantom{+}\alpha\left(|b_2\rangle|00\rangle+|b_5\rangle|01\rangle
+|b_7\rangle|10\rangle+|b_4\rangle|11\rangle\right) \nonumber\\
&&+\beta\left(|b_1\rangle|11\rangle-|b_6\rangle|10\rangle
+|b_8\rangle|01\rangle-|b_2\rangle|00\rangle\right)\nonumber
\end{eqnarray}
where the $|b_j\rangle$ are the 3-particle Bell states defined by
Laflamme {\it et al\/} \cite{Zurek}. Here we have dropped the normalization 
constant. This code is clearly idential to the Laflamme {\it et al\/}
code up to relabelling of the Bell states.

How efficient is this coding scheme? One measure is obtained by asking 
how many `clock cycles' are required to execute the scheme on a quantum 
computer. The most promising device, at least for relatively small numbers 
of two-state systems, is a linear ion-trap model suggested by Cirac and 
Zoller \cite{CZ}. In this model the basic clock frequency is limited by 
that of the center-of-mass mode of the trapped ions as they undergo 
coupled oscillations. This limitation arises from the requirement that 
the laser line-widths be narrower than the lowest vibrational mode of 
the ions, thus ensuring that only the correct energy levels are addressed 
by the laser pulses.  The energy-time uncertainty relation, therefore, 
implies that the duration of the laser pulses must exceed the inverse 
frequency of this lowest vibrational mode. We then conclude that for 
ion-trap computers the number of laser pulses required to complete a 
particular algorithm is a reasonable measure of its efficiency \cite{fn2}.

\begin{figure}[thb]
\begin{psfrags}
\psfrag{p1}[c]{\Large $~~|\psi\rangle$}
\psfrag{z}[c]{\Large $|0\rangle$}
\psfrag{u1}[cb]{\small $~~~~\,\hat U$}
\psfrag{sz}[cb]{$~~~~\,\hat \sigma_z$}
\psfrag{u2}[cb]{\small $~~~~\,\hat U^\dagger$}
\epsfxsize=3.4in
\epsfbox[-20 -10 240 165]{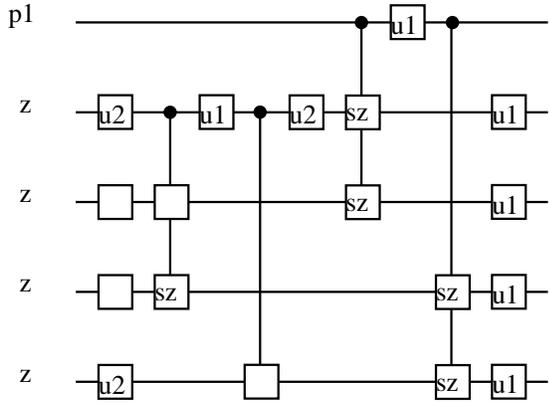}
\end{psfrags}
\caption{Circuit from Fig.~\protect\ref{fig1} rewritten in terms
of the gate-primitives of an ion-trap quantum computer
\protect\cite{CZ}. The single 2-qubit gate is the conditional $\hat\sigma_z$
operation defined in Eq.~(\protect\ref{eq1}) and pairs of them
are drawn as 3-qubit gates. Each single qubit rotation
requires one laser pulse, the 2-qubit gate requires three pulses, and
the 3-qubit gates if implemented as single elements require only four
laser pulses each \protect\cite{Preskill}. This circuit, therefore, uses a
total of 26 laser pulses.}
\label{fig2}
\end{figure}

Rather than directly using the circuit in Fig.~\ref{fig1},
we optimize it for the particular primitive instruction set of the
ion-trap quantum computer in the manner shown in 
Fig.~\ref{fig2}. (Note that the three-qubit operations, the
controlled double $\hat\sigma_z$ operations, require only four
laser pulses each as demonstrated in the appendix.)

A simple counting of the requirements for the circuit of Fig.~\ref{fig2}
yields 26 laser pulses. By contrast the original
circuit of Laflamme {\it et al\/} \cite{Zurek} appears to require at least 
41 laser pulses. Another scheme using six 2-qubit controlled-NOT's and 
five 1-bit gates was mentioned in Ref.~\onlinecite{newpaper}; its
unoptimized form requires 35 laser pulses \cite{DiVincenzo}. 

\begin{figure}[thb]
\begin{psfrags}
\psfrag{p1}[c]{\Large $~~|\psi\rangle$}
\psfrag{z}[c]{\Large $|0\rangle$}
\psfrag{u1}[cb]{\small $~~~~\,\hat U$}
\psfrag{u3}[cb]{\small $~~~~\,\hat V$}
\psfrag{u4}[cb]{\small $~~~~\,\hat W$}
\psfrag{sz}[cb]{$~~~~\,\hat \sigma_z$}
\psfrag{u2}[cb]{\small $~~~~\,\hat U^\dagger$}
\epsfxsize=3.4in
\epsfbox[-20 -10 240 165]{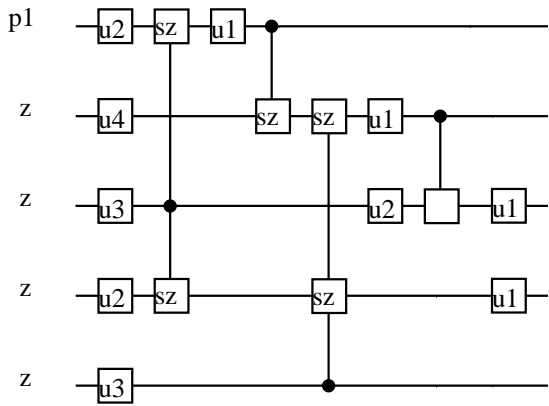}
\end{psfrags}
\caption{The best known circuit for encoding a single qubit in
a five-qubit error correction code. As shown 24 laser pulses would be
required to implement this circuit on an ion-trap quantum computer.}
\label{fig2a}
\end{figure}

It is worth noting here that short of trying all possible circuits
it is not known in general how to determine the optimal circuit.
Much of the optimization achieved here is actually hidden in the
choice of the circuit in Fig.~\ref{fig1} where a number of rearrangements
were tried by hand to yield the `simplest' circuit.
More efficient circuits than shown in Fig.~\ref{fig2} can be found
by computer search. In fact, we show our current best in Fig.~\ref{fig2a},
where we define two new one-bit operations:
\begin{equation}
\hat V = {1\over\sqrt{2}}
\left(\begin{array}{rr}
  1 & -i \\
  -i &  1
\end{array}\right) \;, ~~~~~~
\hat W = \hat V \hat U^\dagger \;.
\end{equation}
As shown it requires only 24 laser pulses, not counting further speedups
such as parallelizing the operation of several of its one-bit gates.

An alternative method of error correction has been suggested by 
Vaidman {\it et al\/} \cite{V}. Its operation invoves a circuit
which can provide only error detection {\it not\/} error correction;
however, by sufficiently rapid operation of the circuit the quantum
Zeno effect allows it to `turn off' the relatively slow errors.
Using the quantum Zeno effect it corrects
for {\it small\/} single-particle perturbations of the system 
rather than the arbitrary single-particle errors of the standard 
schemes. Nonetheless, quantum Zeno error correction has the advantage 
of only requiring 4 qubits. Further, we find that the coding and 
decoding may each be executed using as few as 16 laser pulses with 
possibly only one extra for the auxiliary qubit resetting. How effective
are these error correction schemes that rely on the quantum Zeno
effect? We shall now evaluate their performance for correcting 
phase-diffusion noise.

\section*{Zeno- versus standard quantum-error-correction}

In this section we compare the performance of Zeno and standard 
methods for quantum error correction.  Rather than considering 
the schemes discussed in the previous section, however, we study 
simpler schemes which protect only against 1-qubit dephasing. In 
particular, we compare a compact 2-qubit code given by Chuang and
Laflamme \cite{Ch}, and independently by Vaidman 
{\it et al\/} \cite{V} versus a standard 3-qubit code 
\cite{Steane2,Ekert,ME}. The 2-qubit scheme relies on the quantum 
Zeno effect to correct for {\it small\/} deviations in the system's 
state; whereas the 3-qubit code can correct for arbitrary 1-qubit 
dephasing. How do these schemes compare?

\begin{figure}[thb]
\begin{psfrags}
\psfrag{a}[c]{\Large $|\psi\rangle$}
\psfrag{z}[c]{\Large $|0\rangle$}
\psfrag{e}[c]{${~~~}$1-qubit dephasing}
\psfrag{u1}[cb]{\small $~~~~\hat U$}
\psfrag{u2}[cb]{\small $~~~~\hat U^\dagger$}
\epsfxsize=3.4in
\epsfbox[-30 -20 260 100]{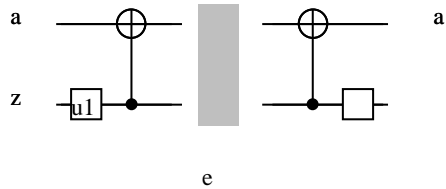}
\end{psfrags}
\caption{Quantum Zeno error correction scheme \protect\cite{Ch,V}. Both 
coding and decoding circuits are shown. (The shaded region represents 
1-qubit dephasing.)}
\label{fig3}
\end{figure}

\begin{figure}[thb]
\begin{psfrags}
\psfrag{a}[c]{\Large $|\psi\rangle$}
\psfrag{z}[c]{\Large $|0\rangle$}
\psfrag{e}[c]{${~~~}$1-qubit dephasing}
\psfrag{u1}[cb]{\small $~~~~\hat U$}
\psfrag{u2}[cb]{\small $~~~~\hat U^\dagger$}
\epsfxsize=3.4in
\epsfbox[-30 -20 260 115]{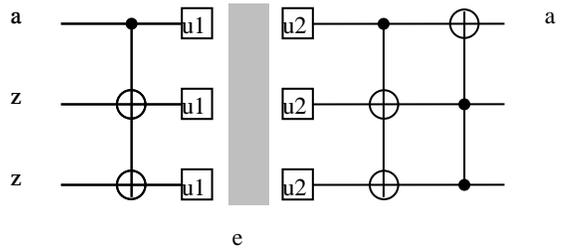}
\end{psfrags}
\caption{Standard quantum 1-bit dephasing correction scheme 
\protect\cite{ME}. Both coding and decoding circuits are shown.}
\label{fig4}
\end{figure}

Figs.~\ref{fig3} and~\ref{fig4} show complete coding and decoding 
circuits for both schemes. Clearly, the Zeno scheme uses fewer resources 
and requires fewer gates to operate so it has a distinct implementational 
advantage over the more conventional schemes.

Our model for dephasing assumes that the phase in each qubit undergoes
an independent random walk according to
\begin{equation}
\alpha |0\rangle +\beta|1\rangle \rightarrow
\alpha |0\rangle +\beta e^{i\phi(t)}|1\rangle \label{model} \;,
\end{equation}
(up to normalization) where the perturbing phases $\phi(t)$
are given by the Ito stochastic calculus \cite{Gardiner} with:
\begin{eqnarray}
\phi(0)& =& 0 \nonumber \\
\langle\!\langle d\phi(t) \rangle\!\rangle &=& 0 \\
\langle\!\langle d\phi(t)\,d\phi(t') \rangle\!\rangle &=& 
2 \delta(t-t')dt \nonumber \;,
\end{eqnarray}
etc., where $d\phi(t)$ is the Ito differential and the doubled angle
brackets represent stochastic averages. Eq.~(\ref{model}) therefore
describes our model of the shaded regions in
Figs.~\ref{fig3} and~\ref{fig4}.

How do each of the above error correction circuits work if applied
only after the dephasing has acted for a time $t$? Delaying the 
decoding circuit in Fig.~\ref{fig3} for a time $t$ after the coding
yields
\begin{equation}
\hat\rho_0 \equiv \left( \begin{array}{cc}
|\alpha|^2 & \overline{\beta}\alpha \nonumber \\
\overline{\alpha}\beta & |\beta|^2 \nonumber
\end{array}\right) \longrightarrow
\left( \begin{array}{cc}
|\alpha|^2 & e^{-t}\, \overline{\beta}\alpha \nonumber \\
e^{-t}\, \overline{\alpha}\beta & |\beta|^2 \nonumber
\end{array}\right) \;,
\end{equation}
here $\hat\rho_0$ is the initial density matrix for the qubit 
$|\psi\rangle$; i.e., there is no improvement using the Zeno error 
correction scheme for this model of noise even for short times!
A similar result was noted by Chuang and Laflamme \cite{Ch}.

By contrast, a delay for time $t$ in circuit~\ref{fig4} before decoding
yields
\begin{eqnarray}
\hat\rho_0 &\longrightarrow& (2+3e^{-t}-e^{-3t})\, \hat\rho_0/4\nonumber \\
&& + (2+ e^{-3t}-3e^{-t})\,\hat\sigma_x \,\hat\rho_0\, \hat\sigma_x /4 \;,
\end{eqnarray}
with $\hat\sigma_x =\left({0\,1\atop 1\,0}\right)$ being one of the 
standard Pauli matrices.

A measure of (relative) coherence between a 
pair of states is given by the absolute value of the off-diagonal 
terms in the density matrix $\hat \rho(t)$ \cite{Damp} 
\begin{equation}
{\cal C}(t)\equiv \left| {\langle 1|\hat\rho(t)|0\rangle \over
\langle 1|\hat\rho_0(t)|0\rangle} \right| \;.
\end{equation}
The 2-qubit Zeno error correction scheme yields
\begin{equation}
{\cal C}_{\rm 2-qubit}(t) = e^{-t} \;,
\end{equation}
whereas the standard 3-qubit scheme has a coherence bounded by its 
worst case 
\begin{equation}
{\cal C}_{\rm 3-qubit}(t) \ge (3 e^{-t}-e^{-3t})/2 \;.
\end{equation}

\begin{figure}[thb]
\begin{psfrags}
\psfrag{c}[c]{\Large ${\cal C}(t)~~$}
\psfrag{t}[c]{\Large $t$}
\epsfxsize=3.4in
\epsfbox[20 180 590 580]{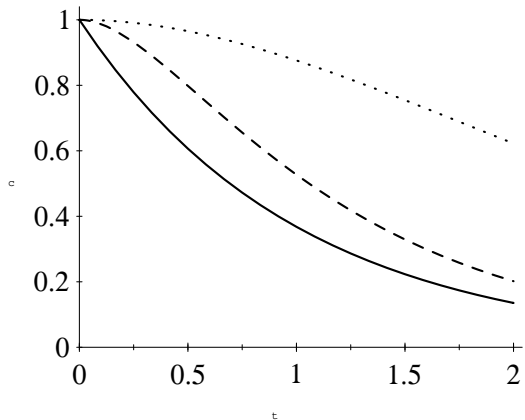}
\end{psfrags}
\caption{Measure of coherence for: no error correction (solid line);
the Zeno error correction scheme of Fig.~\protect\ref{fig3} at time 
$t$ (solid line); the 3-qubit scheme of Fig.~\protect\ref{fig4} at
time $t$ (dashed line, representing the lower bound); and the 10-fold 
repetition of the 3-qubit scheme by time $t$ (dotted line, representing
the lower bound).}
\label{fig5}
\end{figure}

Finally, we note that $n$ evenly spaced repetitions in a time $t$ of an
error correction scheme will yield an improved coherence ${\cal C}$ 
according to
\begin{equation}
{\cal C}^{\,n{\rm-shot}}(t) = \left[ {\cal C}(t/n)\right]^n \;.
\end{equation}
The performance of the Zeno 2-qubit scheme, conventional 3-qubit scheme
and a 10-fold repetition of the latter are displayed in Fig.~\ref{fig5}.

Why does the Zeno error correction scheme fail to work for the noise
model of Eq.~(\ref{model}) even at short times? Put simply, the random
walk model for dephasing implies that the expected deviation of the phase
grows as $\sqrt{t}$ instead of $t$. The error, therefore, accumulates 
too quickly for the Zeno scheme.

The random walk model
of noise really has two time scales: the typical time between random
steps and the much longer dephasing time. The stochastic calculus 
approach assumes that the former of these time scales is so short as to 
be negligible. This means that in this model of noise no matter how 
quickly we operate the Zeno error correction scheme many stochastic 
steps have occurred. The averaging over these many random steps in 
phase produces a pertubabtion that overwhelms the linear correction 
to the state. However, the Zeno error correction schemes discussed
above \cite{Ch,V} require that the change in the system's state be
dominated by linear terms. The implications are that phase 
diffusion is {\it not\/} corrected by these Zeno error correction 
schemes unless they are repeatedly used at a rate faster than the 
typical time between random steps of the phase --- the phase-diffusion 
time itself is already much too slow. How fast does this need to be 
in practice? That depends on the detailed source of the phase 
diffusion: For instance, it might be relatively slow (though still 
faster than the phase diffusion time) when the principle source of 
noise is due to external mechanical noise. Other models, however, are 
very much faster: Unruh's \cite{Unruh} study of decoherence due to 
vacuum fluctuations in the eletromagnetic field coupling to a qubit 
yielded a time-scale comparable to X-ray frequencies.

It is worth mentioning two error `stabilization' schemes which 
utilize the Zeno effect: Zurek \cite{Z2} has outlined a scheme which
averages several copies of a computation, and Berthiaume {\it et al\/}
\cite{B,B2} have considered in some detail a scheme which projects 
several copies of a computation to the symmetric state. Because these 
schemes evenly spread errors over several copies of a computation rather 
than attempt to correct them it may be that they circumvent the 
problem with dephasing discussed here. We leave this question open
for further study.

Quantum error correction of arbitrary single-qubit errors is
rather costly of computing resources: a minimum of five
qubits and possibly 24 laser pulses 
for coding (decoding being only slightly more expensive \cite{fn1}). 
This might be compared with the resources required to execute a moderate 
unprotected calculation; Beckman {\it et al\/} \cite{Preskill} show that 
the Shor algorithm could be implemented on 6 trapped ions using only 
38 laser pulses to factor the number 15 \cite{fn3}. Alternate
error correction schemes based on the quantum Zeno effect are much
more efficient to implement. However, they fail for simple models
of decoherence, such as the model of phase diffusion considered here.

In conclusion, because error correction is virtually as expensive as
the simplest error-correction-free computations, it appears unlikely 
that full quantum error correction will be implemented for computational 
purposes in the first few generations of quantum computers. Instead, 
quantum error correction will probably initially play an important role 
in the long term storage of quantum information: implementing a true 
quantum memory. 

\section*{Appendix: \\
Quantum networks on ion traps}

In this appendix we describe how controlled double $\hat\sigma_z$ 
operations may be performed in four laser pulses on a Cirac-Zoller
ion trap quantum computer \cite{CZ}. These operations are the
{\it three\/}-qubit operations seen in Fig.~\ref{fig2}. Labeling the 
ground and excited states of ion $i$ as $|g\rangle_i$ and 
$|e\rangle_i$ respectively, and the Fock state of the center-of-mass 
vibrational mode of the trap as $|n\rangle_{\rm cm}$ we summarize 
two important operations: A suitably tuned $\pi$-pulse on ion $i$ yields 
the operation \cite{CZ,Preskill}
\begin{equation}
\hat W^{(i)}_{\rm phon} :\left\{
\begin{array}{rcl}
|g\rangle_i |0\rangle_{\rm cm}&\rightarrow&
\phantom{-i}|g\rangle_i |0\rangle_{\rm cm} \\
|g\rangle_i |1\rangle_{\rm cm}&\rightarrow&
-i |e\rangle_i |0\rangle_{\rm cm} \\
|e\rangle_i |0\rangle_{\rm cm}&\rightarrow&
-i |g\rangle_i |1\rangle_{\rm cm} \\
|e\rangle_i |1\rangle_{\rm cm}&\rightarrow&
\phantom{-i}|e\rangle_i |1\rangle_{\rm cm} \;,
\end{array}
\right. 
\end{equation}
whereas, a differently tuned $2\pi$-pulse on ion $j$ yields
\cite{CZ,Preskill}
\begin{equation}
\hat V^{(j)}: \left\{
\begin{array}{rcl}
|g\rangle_j |0\rangle_{\rm cm}&\rightarrow&
\phantom{-}|g\rangle_j |0\rangle_{\rm cm} \\
|g\rangle_j |1\rangle_{\rm cm}&\rightarrow&
-|g\rangle_j |1\rangle_{\rm cm}\\
|e\rangle_j |0\rangle_{\rm cm}&\rightarrow&
\phantom{-}|e\rangle_j |0\rangle_{\rm cm}\\
|e\rangle_j |1\rangle_{\rm cm}&\rightarrow&
\phantom{-}|e\rangle_j |1\rangle_{\rm cm} \;.
\end{array}
\right.
\end{equation}
Finally, another appropriately tuned $\pi$-pulse on ion $j$ yields 
\cite{CZ,Preskill}
\begin{equation}
\hat V^{(j)}_{\rm phon}: \left\{
\begin{array}{rcl}
|g\rangle_j |0\rangle_{\rm cm}&\rightarrow&
\phantom{-i}|g\rangle_j |0\rangle_{\rm cm} \\
|g\rangle_j |1\rangle_{\rm cm}&\rightarrow&
-i|e'\rangle_j |0\rangle_{\rm cm}\\
|e\rangle_j |0\rangle_{\rm cm}&\rightarrow&
\phantom{-i}|e\rangle_j |0\rangle_{\rm cm}\\
|e\rangle_j |1\rangle_{\rm cm}&\rightarrow&
\phantom{-i}|e\rangle_j |1\rangle_{\rm cm} \;,
\end{array}
\right.
\end{equation}
where $|e'\rangle_j$ is a {\it different\/} excited state of ion $j$.

Using these operations and taking the trap's vibrational mode
intially in the ground state $|0\rangle_{\rm cm}$ we find
\begin{eqnarray}
\hat W^{(i)\dagger}_{\rm phon} \hat V^{(k)} 
\hat V^{(j)} \hat W^{(i)}_{\rm phon} :&&
\label{theEq} \\
|\epsilon\rangle_i |\eta_1\rangle_j|\eta_2\rangle_k \rightarrow&&
(-1)^{\eta_1\epsilon}(-1)^{\eta_2\epsilon}
|\epsilon\rangle_i |\eta_1\rangle_j|\eta_2\rangle_k \nonumber \,.
\end{eqnarray}
This completes the construction of the controlled double $\hat\sigma_z$
operation. We note that this construction requires only four laser
pulses as opposed to the six required to perform the two controlled 
$\hat\sigma_z$ operations separately.

In order to see how to generalize this approach let us introduce
a different notation. We start by labeling the states to be
acted on by 
$|\epsilon_1,\,\epsilon_2,\,\ldots,\, \eta_1,\,\eta_2,\,\ldots\rangle$
where the $\epsilon_j$ represents the $j$th control bit and
$\eta_k$ represents the $k$th control bit. When only a single of
either kind of bit occurs we drop the corresponding subscript.
Then we introduce a {\it space-time\/} diagram of events on the
ion-trap to replace the usual circuit notation. 
In these space-time diagrams the horizontal lines represent
the world lines of the ions (in an exactly analogous way that they
do in the usual circuits). Finally, we superpose on these world lines
the events corresponding to an appropriately tuned laser on each ion.
In this way Eq.~(\ref{theEq}) becomes:
\begin{eqnarray}
\begin{picture}(20, 45)
  \unitlength=1pt
  \put(30,40){\circle*{2.5}}
  \put(30,40){\thinlines\line(0,-1){16}}
  \put(30,16){\makebox(0,0){\framebox(15,15)[c]{$\hat\sigma_z$}}}
  \put(30,8.5){\thinlines\line(0,-1){8.5}}
  \put(30,-8){\makebox(0,0){\framebox(15,15)[c]{$\hat\sigma_z$}}}
\end{picture} ~~~~~~~
&\equiv&~~~~~
\begin{picture}(0, 45)
  \unitlength=1pt
  \put(-10,40){\thinlines\line(1,0){10}}
  \put(40,40){\thinlines\line(1,0){30}}
  \put(110,40){\thinlines\line(1,0){10}}
  \put(20,40){\makebox(0,0){$\hat W_{\rm phon}$}}
  \put(90,40){\makebox(0,0){$\hat W_{\rm phon}^\dagger$}}
  \put(-10,16){\thinlines\line(1,0){35}}
  \put(55,16){\thinlines\line(1,0){65}}
  \put(40,16){\makebox(0,0){$\hat V$}}
  \put(-10,-8){\thinlines\line(1,0){55}}
  \put(75,-8){\thinlines\line(1,0){45}}
  \put(60,-8){\makebox(0,0){$\hat V$}}
\end{picture}
\hskip 1.8in \\ \nonumber \\
&=& ~(-i)^\epsilon (-1)^{\bar \eta_1\epsilon} 
(-1)^{\bar\eta_2\epsilon} (+i)^\epsilon \nonumber \\
&=& ~(-1)^{\eta_1\epsilon} (-1)^{\eta_2\epsilon} \nonumber \;,
\end{eqnarray}
where $\bar\eta\equiv(1-\eta)$. Reading from left to right this
circuit decomposes to $\hat W^{(1)\dagger}_{\rm phon} \hat V^{(3)}
\hat V^{(2)} \hat W^{(1)}_{\rm phon}$ where now we must explicitly
add the numbers of the ions. Since these circuits only involve
conditional phase changes it is sufficient to ask what phases
accumulate as we operate the various pulses. We see that whenever
there is an {\it even\/} number of phases to be flipped (i.e.,
an even number of $\hat V$ pulses) that the phases accumulated
from pulses on the control bits are unwanted and need to be cancelled
by applying the inverse operation the second time around.
In particular, here we apply $\hat W_{\rm phon}^\dagger$ the
second time since we applied $\hat W_{\rm phon}$ the first. Similarly,
below where we make use of $\hat V_{\rm phon}$ for a second and 
further control bit we must use $\hat V_{\rm phon}^\dagger$ the
second time whenever there are an even number of bits to have
their phases flipped (i.e., an even number of $\hat V$'s).

We now give two more example constructions:
\begin{eqnarray}
\begin{picture}(20, 95)
  \unitlength=1pt
  \put(30,88){\circle*{2.5}}
  \put(30,64){\circle*{2.5}}
  \put(30,40){\circle*{2.5}}
  \put(30,40){\thinlines\line(0,-1){16}}
  \put(30,40){\thinlines\line(0,+1){48}}
  \put(30,16){\makebox(0,0){\framebox(15,15)[c]{$\hat\sigma_z$}}}
  \put(30,8.5){\thinlines\line(0,-1){8.5}}
  \put(30,-8){\makebox(0,0){\framebox(15,15)[c]{$\hat\sigma_z$}}}
\end{picture} ~~~~~~~
&\equiv&~~~~~
\begin{picture}(0, 45)
  \unitlength=1pt
  \put(-10,88){\thinlines\line(1,0){7.5}}
  \put(15,88){\makebox(0,0){$\hat W_{\rm phon}$}}
  \put(32.5,88){\thinlines\line(1,0){75}}
  \put(125,88){\makebox(0,0){$\hat W_{\rm phon}^\dagger$}}
  \put(142,88){\thinlines\line(1,0){7.5}}
  \put(-10,64){\thinlines\line(1,0){22.5}}
  \put(30,64){\makebox(0,0){$\hat V_{\rm phon}$}}
  \put(47.5,64){\thinlines\line(1,0){45}}
  \put(110,64){\makebox(0,0){$\hat V_{\rm phon}^\dagger$}}
  \put(127.5,64){\thinlines\line(1,0){22}}
  \put(-10,40){\thinlines\line(1,0){37.5}}
  \put(45,40){\makebox(0,0){$\hat V_{\rm phon}$}}
  \put(62.5,40){\thinlines\line(1,0){15}}
  \put(95,40){\makebox(0,0){$\hat V_{\rm phon}^\dagger$}}
  \put(112.5,40){\thinlines\line(1,0){37}}
  \put(-10,16){\thinlines\line(1,0){57.5}}
  \put(60,16){\makebox(0,0){$\hat V$}}
  \put(72.5,16){\thinlines\line(1,0){77}}
  \put(-10,-8){\thinlines\line(1,0){72.5}}
  \put(75,-8){\makebox(0,0){$\hat V$}}
  \put(87.5,-8){\thinlines\line(1,0){62}}
\end{picture}~~~~~
\hskip 1.8in \\ \nonumber \\
&=& ~(-i)^{\epsilon_1} (-i)^{\bar\epsilon_2\epsilon_1}
(-i)^{\bar\epsilon_3\epsilon_2\epsilon_1}
(-1)^{\bar \eta_1\epsilon_3\epsilon_2\epsilon_1} \nonumber \\
&&~~~~ \times(-1)^{\bar\eta_2\epsilon_3\epsilon_2\epsilon_1} 
(+i)^{\bar\epsilon_3\epsilon_2\epsilon_1}
(+i)^{\bar\epsilon_2\epsilon_1} (+i)^{\epsilon_1} \nonumber \\
&=& ~(-1)^{\eta_1\epsilon_3\epsilon_2\epsilon_1}
(-1)^{\eta_2\epsilon_3\epsilon_2\epsilon_1} \nonumber \;,
\end{eqnarray}
which corresponds to the series of laser pulses
\begin{equation}
\hat W_{\rm phon}^{(1)\dagger} \hat V_{\rm phon}^{(2)\dagger}
\hat V_{\rm phon}^{(3)\dagger} \hat V^{(5)} \hat V^{(4)} 
\hat V_{\rm phon}^{(3)} \hat V_{\rm phon}^{(2)} 
\hat W_{\rm phon}^{(1)} \;,
\end{equation}
and as a final example:
\begin{eqnarray}
\begin{picture}(20, 95)
  \unitlength=1pt
  \put(30,88){\circle*{2.5}}
  \put(30,64){\circle*{2.5}}
  \put(30,40){\makebox(0,0){\framebox(15,15)[c]{$\hat\sigma_z$}}}
  \put(30,64){\thinlines\line(0,-1){16}}
  \put(30,64){\thinlines\line(0,+1){24}}
  \put(30,32.5){\thinlines\line(0,-1){8.5}}
  \put(30,16){\makebox(0,0){\framebox(15,15)[c]{$\hat\sigma_z$}}}
  \put(30,8.5){\thinlines\line(0,-1){8.5}}
  \put(30,-8){\makebox(0,0){\framebox(15,15)[c]{$\hat\sigma_z$}}}
\end{picture} ~~~~~~~
&\equiv&~~~~~
\begin{picture}(0, 45)
  \unitlength=1pt
  \put(-10,88){\thinlines\line(1,0){7.5}}
  \put(15,88){\makebox(0,0){$\hat W_{\rm phon}$}}
  \put(32.5,88){\thinlines\line(1,0){55}}
  \put(105,88){\makebox(0,0){$\hat W_{\rm phon}$}}
  \put(122,88){\thinlines\line(1,0){7.5}}
  \put(-10,64){\thinlines\line(1,0){22.5}}
  \put(30,64){\makebox(0,0){$\hat V_{\rm phon}$}}
  \put(47.5,64){\thinlines\line(1,0){25}}
  \put(90,64){\makebox(0,0){$\hat V_{\rm phon}$}}
  \put(107.5,64){\thinlines\line(1,0){22}}
  \put(-10,40){\thinlines\line(1,0){42.5}}
  \put(45,40){\makebox(0,0){$\hat V$}}
  \put(57.5,40){\thinlines\line(1,0){72}}
  \put(-10,16){\thinlines\line(1,0){57.5}}
  \put(60,16){\makebox(0,0){$\hat V$}}
  \put(72.5,16){\thinlines\line(1,0){57}}
  \put(-10,-8){\thinlines\line(1,0){72.5}}
  \put(75,-8){\makebox(0,0){$\hat V$}}
  \put(87.5,-8){\thinlines\line(1,0){44}}
\end{picture}~~~~~
\hskip 1.8in \\ \nonumber \\
&=& ~(-i)^{\epsilon_1} (-i)^{\bar\epsilon_2\epsilon_1}
(-1)^{\bar\eta_1\epsilon_2\epsilon_1} \nonumber \\
&&~~~~ \times (-1)^{\bar\eta_2\epsilon_2\epsilon_1}
(-1)^{\bar\eta_3\epsilon_2\epsilon_1} (-i)^{\bar\epsilon_2\epsilon_1}
(-i)^{\epsilon_1} \nonumber \\
&=& ~(-1)^{\eta_1\epsilon_2\epsilon_1}
(-1)^{\eta_2\epsilon_2\epsilon_1}
(-1)^{\eta_3\epsilon_2\epsilon_1} \nonumber \;,
\end{eqnarray}
which corresponds to the series of laser pulses
\begin{equation}
\hat W_{\rm phon}^{(1)} \hat V_{\rm phon}^{(2)}
\hat V^{(5)} \hat V^{(4)} \hat V^{(3)}
\hat V_{\rm phon}^{(2)} \hat W_{\rm phon}^{(1)} \;.
\end{equation}

\vskip 0.2truein

The authors thank the ISI for giving them the opportunity to
collaborate; SLB appreciated the support of a Humboldt fellowship 
and discussions with N.\ Cohen and D.\ DiVincenzo.


\begin{references}
\bibitem{fn1}Just how many extra operations are required to complete
decoding depends on whether it is done by added circuitry, or by external
observation of the state of the auxiliary qubits. When the necessity 
of re-initializing the auxiliary qubits is included it appears that 
as few as three extra `clock cycles' should be required for the 
external-observation schemes since most of the operations act only 
one single particles and so could probably be executed in parallel.
\bibitem{fn2}Indeed, it is likely that many of the operations
can be run in parallel. For instance, if the block of unitary
operations at the beginning and ending of Fig.~\ref{fig2} are
executed in parallel then the entire circuit would be completed in
only 20 clock cycles.
\bibitem{DiVincenzo}Optimization depends on the detailed structure
of the circuit which was not given in Ref.~\onlinecite{newpaper}.
\bibitem{fn3}The smallest number for which Shor's algorithm strictly
applies is 15.
\end{references}
\end{document}